\documentstyle[twoside]{article}

\catcode`\@=11
\long\def\@makefntext#1{
\protect\noindent \hbox to 3.2pt {\hskip-.9pt  
$^{{\eightrm\@thefnmark}}$\hfil}#1\hfill}               

\def\@makefnmark{\hbox to 0pt{$^{\@thefnmark}$\hss}}    
        
\def\ps@myheadings{\let\@mkboth\@gobbletwo
\def\@oddhead{\hbox{}
\rightmark\hfil\eightrm\thepage}   
\def\@oddfoot{}\def\@evenhead{\eightrm\thepage\hfil
\leftmark\hbox{}}\def\@evenfoot{}
\def\sectionmark##1{}\def\subsectionmark##1{}}

\oddsidemargin=\evensidemargin
\newcounter{sectionc}\newcounter{subsectionc}\newcounter{subsubsectionc}
\renewcommand{\section}[1] {\vspace{12pt}\addtocounter{sectionc}{1} 
\setcounter{subsectionc}{0}\setcounter{subsubsectionc}{0}\noindent 
        {\tenbf\thesectionc. #1}\par\vspace{5pt}}
\renewcommand{\subsection}[1] {\vspace{12pt}\addtocounter{subsectionc}{1} 
        \setcounter{subsubsectionc}{0}\noindent 
        {\bf\thesectionc.\thesubsectionc. {\kern1pt \bfit #1}}\par\vspace{5pt}}
\renewcommand{\subsubsection}[1] {\vspace{12pt}\addtocounter{subsubsectionc}{1}
        \noindent{\tenrm\thesectionc.\thesubsectionc.\thesubsubsectionc.
        {\kern1pt \tenit #1}}\par\vspace{5pt}}
\newcommand{\nonumsection}[1] {\vspace{12pt}\noindent{\tenbf #1}
        \par\vspace{5pt}}

\newcounter{appendixc}
\newcounter{subappendixc}[appendixc]
\newcounter{subsubappendixc}[subappendixc]
\renewcommand{\thesubappendixc}{\Alph{appendixc}.\arabic{subappendixc}}
\renewcommand{\thesubsubappendixc}
        {\Alph{appendixc}.\arabic{subappendixc}.\arabic{subsubappendixc}}

\renewcommand{\appendix}[1] {\vspace{12pt}
        \refstepcounter{appendixc}
        \setcounter{figure}{0}
        \setcounter{table}{0}
        \setcounter{lemma}{0}
        \setcounter{theorem}{0}
        \setcounter{corollary}{0}
        \setcounter{definition}{0}
        \setcounter{equation}{0}
        \renewcommand{\thefigure}{\Alph{appendixc}.\arabic{figure}}
        \renewcommand{\thetable}{\Alph{appendixc}.\arabic{table}}
        \renewcommand{\theappendixc}{\Alph{appendixc}}
        \renewcommand{\thelemma}{\Alph{appendixc}.\arabic{lemma}}
        \renewcommand{\thetheorem}{\Alph{appendixc}.\arabic{theorem}}
        \renewcommand{\thedefinition}{\Alph{appendixc}.\arabic{definition}}
        \renewcommand{\thecorollary}{\Alph{appendixc}.\arabic{corollary}}
        \renewcommand{\theequation}{\Alph{appendixc}.\arabic{equation}}
        \noindent{\tenbf Appendix \theappendixc #1}\par\vspace{5pt}}
\newcommand{\subappendix}[1] {\vspace{12pt}
        \refstepcounter{subappendixc}
        \noindent{\bf Appendix \thesubappendixc. {\kern1pt \bfit #1}}
        \par\vspace{5pt}}
\newcommand{\subsubappendix}[1] {\vspace{12pt}
        \refstepcounter{subsubappendixc}
        \noindent{\rm Appendix \thesubsubappendixc. {\kern1pt \tenit #1}}
        \par\vspace{5pt}}

\newcommand{\textlineskip}{\baselineskip=13pt}
\newcommand{\smalllineskip}{\baselineskip=10pt}

\def\eightcirc{
\begin{picture}(0,0)
\put(4.4,1.8){\circle{6.5}}
\end{picture}}
\def\eightcopyright{\eightcirc\kern2.7pt\hbox{\eightrm c}}

\def\abstracts#1#2#3{{
        \centering{\begin{minipage}{4.5in}\baselineskip=10pt\footnotesize
        \parindent=0pt #1\par 
        \parindent=15pt #2\par
        \parindent=15pt #3
        \end{minipage}}\par}} 

\newcommand{\bibit}{\nineit}

\renewenvironment{thebibliography}[1]
        {\frenchspacing
         \ninerm\baselineskip=11pt
         \begin{list}{\arabic{enumi}.}
        {\usecounter{enumi}\setlength{\parsep}{0pt}     
         \setlength{\leftmargin 12.7pt}{\rightmargin 0pt} 
         \setlength{\itemsep}{0pt} \settowidth
        {\labelwidth}{#1.}\sloppy}}{\end{list}}

\newcounter{itemlistc}
\newcounter{romanlistc}
\newcounter{alphlistc}
\newcounter{arabiclistc}

\newcommand{\fcaption}[1]{
        \refstepcounter{figure}
        \setbox\@tempboxa = \hbox{\footnotesize Fig.~\thefigure. #1}
        \ifdim \wd\@tempboxa > 5in
           {\begin{center}
        \parbox{5in}{\footnotesize\smalllineskip Fig.~\thefigure. #1}
            \end{center}}
        \else
             {\begin{center}
             {\footnotesize Fig.~\thefigure. #1}
              \end{center}}
        \fi}

\newcommand{\tcaption}[1]{
        \refstepcounter{table}
        \setbox\@tempboxa = \hbox{\footnotesize Table~\thetable. #1}
        \ifdim \wd\@tempboxa > 5in
           {\begin{center}
        \parbox{5in}{\footnotesize\smalllineskip Table~\thetable. #1}
            \end{center}}
        \else
             {\begin{center}
             {\footnotesize Table~\thetable. #1}
              \end{center}}
        \fi}

\def\@citex[#1]#2{\if@filesw\immediate\write\@auxout
        {\string\citation{#2}}\fi
\def\@citea{}\@cite{\@for\@citeb:=#2\do
        {\@citea\def\@citea{,}\@ifundefined
        {b@\@citeb}{{\bf ?}\@warning
        {Citation `\@citeb' on page \thepage \space undefined}}
        {\csname b@\@citeb\endcsname}}}{#1}}

\newif\if@cghi
\def\cite{\@cghitrue\@ifnextchar [{\@tempswatrue
        \@citex}{\@tempswafalse\@citex[]}}
\def\citelow{\@cghifalse\@ifnextchar [{\@tempswatrue
        \@citex}{\@tempswafalse\@citex[]}}
\def\@cite#1#2{{$\null^{#1}$\if@tempswa\typeout
        {IJCGA warning: optional citation argument 
        ignored: `#2'} \fi}}

\def\pmb#1{\setbox0=\hbox{#1}
        \kern-.025em\copy0\kern-\wd0
        \kern.05em\copy0\kern-\wd0
        \kern-.025em\raise.0433em\box0}

\def\fnt#1#2{\footnotetext{\kern-.3em
        {$^{\mbox{\scriptsize #1}}$}{#2}}}

\font\tenrm=cmr10
\font\tenit=cmti10 
\font\tenbf=cmbx10
\font\bfit=cmbxti10 at 10pt
\font\ninerm=cmr9
\font\nineit=cmti9

\font\eightrm=cmr8

\def\qed{\hbox{${\vcenter{\vbox{                        
   \hrule height 0.4pt\hbox{\vrule width 0.4pt height 6pt
   \kern5pt\vrule width 0.4pt}\hrule height 0.4pt}}}$}}

\begin{document}

\newcommand{\be}{\begin{equation}}
\newcommand{\ee}{\end{equation}}
\newcommand{\ben}{\begin{eqnarray}}
\newcommand{\een}{\end{eqnarray}}
\newcommand{\n}{\label}
\newcommand{\no}{\noindent}

\centerline{\bf STABLE INFLATIONARY DISSIPATIVE COSMOLOGIES}
\vspace*{0.37truein}
\centerline{\footnotesize LUIS P. CHIMENTO\footnote{Electronic address:
chimento@df.uba.ar} and ALEJANDRO S. JAKUBI\footnote{Electronic address:
jakubi@df.uba.ar}}
\centerline{\footnotesize Departamento de F\'{\i}sica, Universidad
de Buenos Aires}
\baselineskip=10pt
\centerline{\footnotesize 1428 Buenos Aires, Argentina}
\vspace*{10pt}
\centerline{\footnotesize DIEGO PAVON\footnote{Electronic address: 
diego@ulises.uab.es}}
\vspace*{0.015truein}
\centerline{\footnotesize Departament de F\'{\i}sica, Universidad Aut\'onoma 
de Barcelona}
\baselineskip=10pt
\centerline{\footnotesize 08193 Bellaterra, Spain}
\vspace*{0.21truein}

\abstracts
{The stability of the de Sitter era of cosmic expansion in spatially 
curved homogeneous isotropic universes is studied. The source of
the gravitational field is an imperfect fluid such that the parameters
that characterize it may change with time. In this way we extend
our previous analysis for spatially-flat spaces as well
as the work of Barrow.}{}{}

\vspace*{1pt}\textlineskip
\section{Introduction}
\vspace*{-0.5pt}
\noindent
Inflationary cosmic expansions driven by a dissipative fluid has attracted
some attention in the past as a mechanism able to solve the problems usually
attached to the standard hot big-bang scenario (i.e. flatness, horizons, and
monopoles abundance) different from those based on one (or more) scalar
fields. The main dynamic effect of either the scalar field(s) or the
dissipative fluid is to produce a state of cosmic accelerated expansion
through the violation of the strong energy condition. If this accelerated
state is sufficiently long, and if the Universe is able to exit it, the
mentioned problems may find a satisfactory solution. Very often this analysis
has been restricted to spatially-flat
Friedmann-Lema\^{\i}tre-Robertson-Walker (FLRW for short) space-times, if
only because the dependence of the scale factor on the spatial curvature $k$
becomes negligible shortly after inflation has set in. However, one should be
mindful that this curvature may have some impact on the stability of the
inflationary solutions. Given the potential importance of such a scenario, it
is advisable to stablish its generality with respect to the effect of spatial
curvature. In \cite{lad} we studied the stability of inflationary expansions
caused by a dissipative fluid governed by a causal transport equation (i.e.
one that respects both relativistic causality and hydrodynamic stability of
the fluid \cite{israel}, \cite{djc}), assuming $k = 0$. Before going any
further, a word of caution seems in order. The stability of the dissipative
fluid, in the sense this expression is used here, depends on the physical
properties of the fluid under consideration, i.e. on its state and transport
equations \cite{hl1}, \cite{hl2}. Accordingly it should not be confused with
the stability of the cosmic expansion -either inflationary or not. The
stability of inflationary expansions driven by dissipative fluids in curved
FLRW space-times was briefly discussed by Barrow \cite{barrow1},
\cite{barrow2}. However, his study was confined to fluids governed by
transport equations of Eckart's type \cite{eckart} which, as is well-known,
entails the aforesaid problems of acausality and instability.

The aim of this paper is to investigate the combined effects of viscosity 
and curvature
extending the
analysis of \cite{lad} to spatially curved FLRW space-times
and to study scenarios not considered previously.
This bears some interest because observationally it is
doubtful whether the spatial curvature of the Universe is
positive, null or negative. On the one hand the average 
matter density is well below the critical value 
($\sim 0.3$ in critical units), but on the 
other hand the Universe seems to have entered an accelerated phase 
nowadays, which strongly hints to a cosmological constant of
about $0.7$ in the same units -however its value is limited
by gravitational lensing \cite{bartelmann}. Further, although 
the location of  the first acoustic peak in the CBR spectrum 
seems to suggest a flat Universe, its exact position is still
uncertain \cite{silk}, \cite{coble}.
In extending the results of \cite{lad}
we also extend the work of Barrow's since the transport
equation we consider is of causal type and therefore more general
than the one used by him -for a brief illustration of the 
impact of causal transport equations in FLRW cosmology 
see \cite{pbj} \cite{visco} \cite{roy}  \cite{winfried} and \cite{Chi96d}.

To analyse the asymptotic stability of the inflationary 
expansions we will resort again to the second method of Lyapunov 
\cite{Ces}, succintly stated in the appendix of \cite{lad}.  
In addition, some of the scenarios considered in this paper 
will be similar to those of \cite{lad}, whereby we shall 
avoid unnecessary repetition of details. As it turns out, the
set of de Sitter solutions is drastically restricted because of
a constraint that links some of the fluid parameters with 
the rate of cosmic expansion.
\noindent
Our choice of units is $c = 8\pi G = k_{B} = 1$.

\section{Asymptotic Stability}
\noindent
We consider a FLRW universe filled with a imperfect fluid whose
dissipative bulk pressure obeys the causal transport equation 
\cite{israel}, \cite{djc}, \cite{roy}, \cite{winfried}
\begin{equation}
\pi + \tau\dot{\pi}  =  - 3\zeta H 
\label{dpi2}
\end{equation}
where $\tau (> 0)$ is the relaxation time associated at the dissipative
process, $\zeta (> 0)$ the phenomenological coefficient of bulk
viscosity, $H \equiv \dot{a}(t)/a(t)$  denotes the Hubble function,
and $a(t)$ the scale factor of the FLRW metric. The corresponding
Einstein field equations are 
\begin{equation}
H^{2} = \frac{1}{3} \rho - \frac{k}{a^{2}}  \qquad (k = 0, \pm 1)
\label{00}
\end{equation}
\begin{equation}
3 (\dot{H} + H^{2}) = - \frac{1}{2} \left[\rho + 3(p+\pi)\right]
\label{000}
\end{equation}
\noindent
with $\rho$ and $p$ the energy density and hydrostatic pressure,
respectively. These latter two quantities are assumed to be linked
by the equation of state $p = (\gamma - 1) \rho$, where $\gamma$
denotes the adiabatic index of the cosmic fluid. This, in general,
depends on time -though oftenly, for simplicity, it is assumed 
constant.

\noindent
The above set of equations yields
$$
\ddot{H}+3\gamma H\dot{H}+\tau^{-1}\left[\dot{H}+{\textstyle{3\over
2}}\left(\gamma+\tau\dot\gamma\right)H^{2} -
{\textstyle{3\over 2}}\,\zeta H\right]
$$
\begin{equation}
+\frac{k}{a^{2}} \left[(1- {\textstyle{3\over
2}}\gamma)\left(2H-\tau^{-1}\right)
+ {\textstyle{3\over 2}} \dot{\gamma}\right]=0
\label{5}
\end{equation}

\noindent
and the latter can be recast as

\begin{equation} \label{d2}
\frac{d}{dt}\left[\frac{1}{2}{\dot H}^2+V(H)\right]=D(H,\dot H).
\end{equation}
\noindent
Here, the left hand side is the time derivative of a Lyapunov function
(see appendix in \cite{lad}) with

$$
V(H)=\frac{1}{2}
\left(\gamma+\tau\dot\gamma\right)\tau^{-1}H^3+
\left[\frac{k}{a^2}\left(1-\frac{3}{2}\gamma\right)
-\frac{3}{4}\zeta\tau^{-1}\right]H^2
$$

\begin{equation} \label{V}
+\frac{k}{a^2}\left[\frac{3}{2}\dot\gamma-
\left(1-\frac{3}{2}\gamma\right)\tau^{-1}\right]H
\end{equation}

\noindent
and

$$
D(H,\dot H)=-\left(3\gamma H+\tau^{-1}\right)\dot H^2+
\frac{1}{2}\tau^{-2}\left(\tau\dot\gamma-\dot\tau\gamma+
\tau^2\ddot\gamma\right)H^3
$$

$$
+\frac{3}{4}\tau^{-2}\left(\zeta\dot\tau-\dot\zeta\tau\right)H^2
-\frac{2k}{a^2}\left(1-\frac{3}{2}\gamma\right)H^3
-\frac{3k}{2a^2}\dot\gamma H^2
$$

\begin{equation} \label{D}
-\frac{2k}{a^2}\left[\frac{3}{2}\dot\gamma
-\left(1-\frac{3}{2}\gamma\right)\tau^{-1}\right]H^2
+\frac{k}{a^2}\left[\frac{3}{2}\ddot\gamma+\frac{3}{2}\dot\gamma\tau^{-1}+
\left(1-\frac{3}{2}\gamma\right)\tau^{-2}\dot\tau\right]H.
\end{equation}

\noindent
As shown in figure 1 the potential (\ref{V}) has two extrema, 
a minimum at $H^+$ and a maximum at $H^-$. Their asymptotic 
behaviours for an expanding evolution at late time are

\begin{equation} \label{Hp}
H^+=\frac{\zeta}{\gamma+\tau\dot\gamma}+{\cal O}(a^{-2})
\end{equation}
\noindent
and
\begin{equation} \label{Hm}
H^-=\frac{k}{9\zeta a^2}\left[\frac{3}{2}\dot\gamma-
\left(1-\frac{3}{2}\gamma\right)\tau^{-1}\right]+{\cal O}(a^{-4}).
\end{equation}

\noindent The solution (\ref{Hm}) is unstable, while the stability of the
solution (\ref{Hp}) is determined by the sign of the leading term of $D$ in a
neighbourhood of the point $(H_0,0)$ of the phase plane $(H,\dot H)$, where
$H_{0}$ is the leading term  of $H^{+}$

\begin{equation} \label{D0}
D\approx -\left(3\gamma H+\tau^{-1}\right)\dot H^{2}+
\frac{1}{3}H^2\left(H-\frac{3}{2}H^+\right)
\frac{d}{dt}\left(3\gamma H+\tau^{-1}\right)+{\cal O}(a^{-2}).
\end{equation}
\noindent 
Here we have assumed that $H^{+}$ is a quasi-de Sitter solution.
A sufficient condition for this solution to be asymptotically stable
is that $3\gamma H+\tau^{-1}$ increases for large time. This is equivalent to
the condition that $\zeta/\tau$ increases with time.
Then using the relationship \cite{roy}
\begin{equation}
\frac{\zeta}{\tau} = v^2\gamma\rho
\label{zeta}
\end{equation}
\noindent 
where $v$ is speed of the dissipative signal associated to $\pi$, 
this condition is also equivalent that $\gamma v^2$ is an increasing 
function of time.

It can be seen that $H_0$ is an exact de
Sitter solution when the constraint
\begin{equation} \label{con}
\dot\gamma+\left[\tau^{-1}-2H_0\right]\gamma=\frac{2}{3}
\left[\tau^{-1}-2H_0\right]
\end{equation}
\noindent
is satisfied. It gives for $\gamma\neq 2/3$

\begin{equation}
\label{H0}
H_{0}=\frac{-1\pm\sqrt{1+6\left(3\gamma-2\right)\tau\zeta}}
{2\left(3\gamma-2\right)\tau}
\end{equation}

\noindent
and

\begin{equation} \label{H02/3}
H_{0}=\frac{3}{2}\zeta
\end{equation}

\noindent when $\gamma=2/3$. Also the static solution $H=0$ exists for $k=1$,
when the constraint

\begin{equation} \label{con0}
\tau\dot\gamma+\gamma=\frac{2}{3}
\end{equation}

\noindent holds. To analyse its stability we linearize 
equation (\ref{5}) about it,

\begin{equation} \label{delta}
\ddot H+\tau^{-1}\dot H+3\left[\frac{\dot\gamma\tau}{a^2}
-\frac{1}{2}\zeta\tau^{-1}\right]H=0.
\end{equation}

\noindent The general solution of (\ref{delta}) grows for any decreasing
adiabatic index showing that the static solution is unstable and can be
indentified with the limiting case of the unstable solution $H_{-}$. 
This result could be different if a non-standard increasing adiabatic
index in the limit $t\to\infty$ were admitted. However, by fine tuning 
the initial condition a one-parameter family of solutions that approaches 
a static spacetime at large times can be obtained. 
Besides, there exists a one-parameter family of solutions that starts 
from a static spacetime in the far past and evolves towards a stable 
de Sitter solution.

\section{Inflationary Solutions}

\noindent
Accelerated expansion requires a sufficiently negative pressure, hence we will
consider evolutions that approach asymptotically to a negatively constant
viscous pressure. To this end we will investigate three models, two of them
have the viscous pressure coefficient proportional to Hubble time, and in the
last one it is determined by causality and stability of scalar perturbations.
These three cases are characterized by the specific instances:
(i) when $\dot{\gamma} = 0$ and $ \zeta \propto H^{-1}$, 
(ii) when $\tau$ and $\zeta$ vary as $H^{-1}$, and 
(iii) when $\tau^{-1}\propto H\, $ and $\, \zeta=v^2\gamma\rho\tau$. We
will see them in turn.

\subsection{Constant adiabatic coefficient model}
When $\zeta = $ constant (\ref{5}) can be rewritten as
$$
\tau\frac{d}{dt}\left[\dot H+\frac{3}{2}\gamma H^2-
\frac{k}{a^2}\left(1-\frac{3}{2}\gamma\right)\right]
$$
\begin{equation} \label{dH2}
+\left[\dot H+\frac{3}{2}\gamma H^2-
\frac{k}{a^2}\left(1-\frac{3}{2}\gamma\right)-
\frac{3}{2}\zeta H\right]=0
\end{equation}
\noindent
accordingly, the bulk viscosity coefficient must be  given by
\begin{equation} \label{zeta1}
\zeta=\frac{2c}{3 H}
\end{equation}
\noindent
with $c$ a positive constant, and
\begin{equation} \label{dH3}
\dot H+\frac{3}{2}\gamma H^2-
\frac{k}{a^2}\left(1-\frac{3}{2}\gamma\right)=c
\end{equation}
\noindent 
is a first integral of (\ref{dH2}).
Assuming that the adiabatic index does not vary with time, this
equation can be easily reduced to an equivalent mechanical system by the
substitution $a=s^{2/3\gamma}$ (in the appendix we obtain its parametric
general solution), with first integral

\begin{equation} \label{ds}
\frac{\dot s^2}{2}+\frac{9\gamma^2 k}{8}s^{2\left(1-2/3\gamma\right)}-
\frac{3\gamma c}{4}s^2=\frac{3\gamma E}{2}
\end{equation}

\noindent where $E$ plays the role of a mechanical energy. This expression
shows that the curvature and viscosity terms behave as ``conservative
forces". In particular viscosity provides an effective spring with negative
elastic constant, and it determines the late time behaviour of this model. 
Next we will integrate this system for some values of $\gamma$ that yield 
simple solutions and were considered in \cite{barrow1} to compare with the 
results obtained for fluids governed by  Eckart's transport equation:

I) When  $\gamma=1/3$ we obtain

\begin{equation} \label{at1}
a(t)=c_1+c_2 \exp\left(\sqrt{2c}t\right)+c_3 \exp\left(-\sqrt{2c}t\right)
\end{equation}
\noindent
provided the constraint $8cc_2c_3-k-2cc_1^2=0$ holds. In the limit $c\to 0$
the usual perfect fluid solutions are recovered

\begin{equation} \label{at2}
a(t)=a_{1}t^{2}+a_{2} t+a_{3}
\end{equation}

\noindent
provided the constraint $4a_1a_3-a_2^2-k=0$ is satisfied. This 
family of solutions includes several interesting possibilities.
Solutions with and without initial singularity, bouncing solutions 
and solutions with a finite
time-span, as well as the Milne solution when $k=-1$ and $a_{1}=0$.

II) When $\gamma=2/3$ there are three families of solutions depending
on the sign of $E$:
\begin{enumerate}
\item For $E> 0$
\begin{equation} \label{at23+}
a(t)=\sqrt{\frac{2E}{c}}\sinh\left(\sqrt{c}\Delta t\right)
\end{equation}
\no i.e. the scale factor initially grows as $\sqrt{2E}\Delta t$
independently of $c$.
\item For $E<0$
\begin{equation} \label{at23-}
a(t)=\sqrt{-\frac{2E}{c}}\cosh\left(\sqrt{c}\Delta t\right)
\end{equation}

\no this family of non-singular solutions bounce at $\Delta t=0$ 
when the scale factor attains its minimum value $a_m=\sqrt{-2E/c}$.

\item For $E = 0 $
\begin{equation} \label{at230}
a(t)=\sqrt{\frac{1}{c}}\exp\left(\sqrt{c}
\Delta t\right).
\end{equation}
\end{enumerate}
\no The three families of solutions exist for all $k=0$ or $\pm 1$, 
and for large time they
approach the inflationary de Sitter scenario $a(t)\approx
\exp\sqrt{c}\Delta t$. These results extend those of \S4
in \cite{barrow1} obtained in the framework of the 
non--causal Eckart's theory,
and where the viscosity was assumed to follow a power-law
dependence upon the density.

III) When $\gamma=4/3$ (i.e. a radiation-dominated universe) 
one has
\begin{equation} \label{at43}
a^2(t)=\frac{1}{c}\left[k+b\exp\left(-\sqrt{2c}t\right)
+\frac{1}{8b}\left(1-4cE\right)\exp\left(\sqrt{2c}t\right)\right]
\end{equation}

\noindent
where $b$ is an integration constant.
This familiy of solutions comprises three types of behaviours depending 
on the value of $E$. When $E>0$ singular solutions occur for all $k$.
However, when $-1/4c<E<0$ a singularity arises only if $k<0$. Near 
the singularity the behaviour of the solution (\ref{at43}) is 
radiation--like. When $E<-c$ there is no singularity.

Again the large-time behaviour is qualitatively the same as that
found with Eckart's theory. Otherwise the evolution behaves differently.

IV) When $\gamma>2/3$ it follows, using the results of the appendix,
a family of solutions that near the
singularity evolves as $a\approx t^{2/3\gamma}$, and  at late time it 
exhibits a de Sitter expansion $a\approx \exp \sqrt{2c/3\gamma}t$,
irrespective of the spatial curvature.

\subsection{Constant interactions number per expansion time model}

The expanding universe defines a natural time--scale  -- the expansion time
$H^{-1}$. Any particle species will remain in thermal equilibrium
with the cosmic fluid so long as the interaction rate is high enough to allow
rapid adjustment to the falling temperature. The relaxation time $\tau$ is the
characteristic colisional time of hydrodynamical procesess occurring after the
quantum era. Then a necessary condition for maintaining thermal equilibrium is

\begin{equation}
\tau < H^{-1}
\label{1.}
\end{equation}

\noindent
Now $\tau$ is determined by

\begin{equation}
\tau\simeq{1\over n\sigma v}
\label{2.}
\end{equation}

\noindent 
where $\sigma$ is the interaction cross--section, $n$ is the number
density of the target particles with which the given species is interacting,
and $v$ is the mean relative speed of interacting particles. So we have that
$\nu=\left(\tau H\right)^{-1}$ is the number of interactions in an expansion
time. Now we will consider the case when $\nu$ is a constant larger than one.

To find the solutions of equation (\ref{5}) that satisfy this property we
insert the ansatz

\begin{equation} \label{gamma4}
\gamma=\frac{2}{3}\left(1+\epsilon\right)
\end{equation}

\noindent along with (\ref{zeta1}) in the equation (\ref{5}).
A set of solutions can be found for a variable adiabatic index.
Expression (\ref{5}) splits in two equations,

\begin{equation} \label{dHc}
\ddot H+(2+\nu) H\dot H+\nu H^3-\nu c H=0
\end{equation}

\noindent
plus a linear equation in $\epsilon$,

\begin{equation}
\n{epsilon}
\left[H^2+\frac{k}{a^2}\right]\dot\epsilon+\left[2H\left(\dot H-
\frac{k}{a^2}\right)+\tau^{-1}\left(H^2+\frac{k}{a^2}\right)\right]\epsilon=0
\end{equation}

\no provided $\tau^{-1}=\nu H$ with $\nu$ a constant. 
Solving (\ref{epsilon}), we get

\begin{equation} \label{gamma5}
\gamma=\frac{2}{3}\left(1+b\frac{a^{2-\nu}}{k+\dot a^2}\right)
\end{equation}

\noindent where $b$ is a free parameter. To make explicit the $t$ dependence
of the adiabatic index we need solutions of (\ref{dHc}) that satisfy
$0\le\gamma\le 2$. Equation (\ref{dHc}) can be transformed into a second order
linear differential equation whose general solution, obtained in \cite{luis},
reads
\begin{equation}
\n{Hsol}
H^2=c+\frac{c_1}{a^2}+\frac{c_2}{a^{\nu}}
\end{equation}
\noindent 
where $c_1$ and $c_2$ are arbitrary integration constants. There it was
shown that simple explicit solutions of (\ref{dHc}) can be obtained when
$\nu=1$ and when $\nu=4$. In this case the solutions
are
\begin{equation}
\n{1}
a(t)=c_1+c_2 \exp\left(\sqrt{c}t\right)+c_3 \exp\left(-\sqrt{c}t\right)
\end{equation}
and
\begin{equation}
\n{4}
a(t)=\left[c_1+c_2 \exp\left(2\sqrt{c}t\right)+
c_3 \exp\left(-2\sqrt{c}t\right)\right]^{1/2}
\end{equation}

\noindent 
respectively. For $t \gg c^{-1/2}$, both (\ref{1}) and (\ref{4})
describe inflationary expansions regardless the initial 
conditions. 

\subsection{Stable causal sound perturbations model}
\noindent
In this subsection we will investigate the consequences imposed by causality
and stability on scalar longitudinal sound perturbations. It can be seen that
the relationship between the viscosity coefficient and the dissipative
contribution to the sound speed is $\zeta=v^2\gamma\rho\tau$.

In this case equation (\ref{5}) transforms into

\begin{equation} \label{dh}
h''+\left(3\gamma+\nu\right)h'+3\gamma\left(\nu-3v^2\right)h=0
\end{equation}
\noindent 
where the variable $h=H^2+k/a^2$ is proportional to the energy
density of the fluid and the prime indicates derivative with respect to
$\eta=\ln a$. Except for  $\nu = 3v^2$, the general solution of this equation
leads  asymptotically to power--law behaviors \cite{visco}\cite{Ale}. 
When  $\nu = 3v^2$ the solution exhibit de Sitter expansion at late time. 
Exact solutions of (\ref{dh}) are

\begin{equation} \label{atv-1}
a(t)=\frac{1}{\sqrt{b}}\sinh\left(\sqrt{b}\Delta t\right),
\qquad (k=-1)
\end{equation}

\begin{equation} \label{atv+1}
a(t)=\frac{1}{\sqrt{b}}\cosh\left(\sqrt{b}\Delta t\right),
\qquad (k=1)
\end{equation}
\noindent 
where $b$ is an integration constant. These solutions are 
attractors and therefore of singular importance as they indicate 
the leading behaviour for large cosmological times. It is
remarkable that the expressions (\ref{atv-1}) and (\ref{atv+1}) do 
not explicitly contain any quantity directly associated to viscosity 
(or particle creation) even though it is precisely this effect that 
drives the exponential expansion. In this case the same effect
generates the translational invariance of the energy density.

\section{Additional Inflationary Scenarios}
\subsection{Variable $\tau$}
\noindent
When either $\gamma$ or $\zeta$ is a constant the solution 
(\ref{H0}) is asymptotically
stable provided $\tau$ is a decreasing function. When $\tau$
is a function that first decreases and in a second stage 
increases, there is a first period of exponential inflation
followed by a graceful exit.                   
On the other hand, when the relationship (\ref{zeta}) holds,
$\tau$ presents a minimum provided $v^{2}$ has a
maximum. In this case the Universe enters an inflationary stage 
and afterwards exits it. This parallels the cosmic scenario of 
\S$2.2$ in \cite{lad}. There the cosmic fluid was modelled as a 
mixture of radiation  and heavy particles 
that decayed at a
very high rate into (more stable) lighter particles 
(less massive modes), with high or moderate multiplicity 
\cite{wdr}. 

\subsection{Variable $\tau$ and $\gamma$}
\noindent
Another model of dissipatively driven inflation arises
when $\zeta$ is a constant (or nearly a constant) and $\gamma $ 
varies as
\begin{equation} \label{gammat2}
\gamma(t)=\frac{2}{3}\left[1+\frac{c}{\tau(t)}\right]
\end{equation}
\noindent 
with  $c$ a constant. From (\ref{D}) we see that $D$ is 
negative-definite when $\dot\tau<0$, and accordingly 
the de Sitter expansion results asymptotically stable. 
Likewise $\gamma(t)$ increases when $c>0$. 

By choosing  $c=\tau(t_1)/2$ and $\tau(t_1)/\tau(t_2)=2$
one follows a model in which initially all the energy is in 
the form of non-relativistic particles, $\gamma = 1$, and it
is gradually transferred to relativistic ones, so that finally
$\gamma = 4/3$. This corresponds to the dissipative 
process of decay of dust particles into radiation. Additional 
dissipation may arise from the interaction matter-radiation.
Once the heavy particles have decayed, the Universe exits 
inflation. 
A similar scenario in flat space was reported in \cite{lad}, 
(see equations (17) \& (18) there) but in that work $\tau$ 
was considered constant instead.   

Assuming that the dust particles are primeval mini-black
holes of rest mass $m$ and that the thermodynamic
properties of the mixture of black holes
and relativistic particles correspond to 
a Boltzmann gas of vanishing chemical potential,
the adiabatic index reads
\begin{equation} \label{gammabeta}
\gamma(z)=1+\frac{K_{2}(z)}{z K_{1}(z)+3K_2(z)}
\end{equation}
\noindent
where $K_{n}$ are modified Bessel functions of the second kind,
and $z \equiv m/T$ the dimensionless inverse temperature.
By equating the right hand sides of (\ref{gammat2}) and (\ref{gammabeta})
we can describe the continuous process of decay of mini-black holes 
from $t = t_{1}$, when the black hole energy density dominates
the Universe, until $t_{2}$ when the mini-black holes have
completely evaporated away and the Universe becomes
radiation--dominated. Here it is understood
that all the black holes have the same mass and therefore the same
temperature, and that this one  equals the temperature of the 
massless component of the cosmic fluid at the beginning of the
evaporation. The temperature behaves as in the
$k = 0 $ case \cite{lad} 
\begin{equation} \label{dz}
\frac{z'}{z}=\frac{12 K_2(z)+3z K_1(z)-B z^2}{12 K_2(z)+5z K_1(z)+z^2 K_0(z)}
\end{equation}
\noindent 
where $B=9H\zeta/A_0$, with $A_{0}$ a constant and $'\equiv d/Hdt$. From
this equation follows that $z'$ is negative for
large $z$, which agrees with the warm inflationary scenario of above.

The time dependence of this temperature near $t_{1}$ and $t_{2}$ 
for a generic $\tau(t)$ can be made explicit by expanding
(\ref{gammat2}) about $t = t_{1}$ and 
(\ref{gammabeta}) for $z \to \infty$. Thus one follows
$T \propto t-t_{1}$, while  in the opposite limit 
(i.e. $t\to t_{2}$ and $z \to 0$) it yields 
$T \propto (t_{2}-t)^{-1/2}$. This reproduces the 
results of \cite{lad} but this time with a generic $\tau(t)$.
Again the  production of relativistic particles at the final stage of 
the black holes evaporation is accompanied by  a huge increase 
of the temperature of the cosmic fluid -which also agrees with a 
previous study of this process \cite{wzdp}.
 
The entropy production per unit volume in the radiation fluid
is given by the well-known expression \cite{israel}, \cite{djc}, 
\cite{roy},
$\dot S= \pi^{2}/(\zeta T)$. It has the limiting behaviours
\begin{equation} \label{dS1}
\dot S \simeq -
\frac{3\tau(t_1)\pi^{2}(t_{1})}{m\zeta\dot\tau(t_{1})
\left(t-t_{1}\right)}, \qquad t\to t_{1}
\end{equation}

\begin{equation} \label{dS2}
\dot S\simeq 2 \frac{\pi^{2}(t_{2})\sqrt{6\dot\tau(t_2)\left(t-t_2\right)}}
{m\zeta},
\qquad t\to t_2
\end{equation}

\noindent
where

$$
\pi^{2}(t) = \frac{1}{\tau^{2}(t)}\left\{\frac{1}{2\tau^{2}(t_1)}
\left(1+\frac{2\tau(t)}{\tau(t_1)}\right)^{2}
\left[3\tau(t_1)\zeta\left(
2+\frac{3}{2}\tau(t_1)\zeta\right)+\right.\right.
$$
$$
\left. 1-\left(1+3\tau(t_1)\zeta\right)
\sqrt{1+6\tau(t_1)\zeta}+1\right]+
$$
\begin{equation} \label{pi2}
\left.\frac{k}{a^{2}(t)}\left(1+\frac{2\tau(t)}{\tau(t_1)}\right)\left(1+
3\tau(t_1)\zeta-\sqrt{1+6\tau(t_1)\zeta}\right)+
\frac{\tau^{2}(t_{1})k^2}{a^{4}(t)}
\right\}
\end{equation}

\noindent
and

\begin{equation} \label{at}
a(t)=a_0\exp\left\{\left[
\frac{\sqrt{1+6\zeta\tau(t_1)}-1}{2\tau(t_1)}\right]t\right\}.
\end{equation}
The entropy production rate in the radiation fluid happens to
be very high at the beginning of the evaporation, but 
decreases sharply about the end of this process. As in \cite{lad} 
the net rate of radiation particle production per
mini-black hole and unit of volume varies roughly as 
$(\rho + p)^{-1}$, where in this case $\rho $ and $p $ refer
to the radiation fluid only \cite{wzdp1}. 

\section{General case}
\noindent
In this section we consider the de Sitter solution (\ref{H0}) and assume 
that $\gamma, \tau$ and $\zeta$ vary with time. Again this more
general situation may occur during the decay of massive particles into lighter
ones and also during the decay of four-dimensional fundamental strings into
massive and massless particles -admittedly this second possibility is more
speculative. In this case the algebraic relationship holds

\begin{equation} \label{gammat3}
\gamma(t)=\frac{2}{3}+\frac{3\zeta(t)-2H_0}{6H_0^2\tau(t)}.
\end{equation}
\noindent
By adequately choosing the behaviour of $\zeta(t)$ and $\tau(t)$ 
the decay into radiation, reaching relativistic gas state when 
$\gamma=4/3$, can be described.
After the decay a condensation phase back into non-relativistic 
matter may occur. It ends when $\gamma$ returns to $1$. A 
scenario compatible with the latter phase is the quantum tunneling 
of radiation into black holes \cite{gpy}. This may
arise very naturally because of the instability of the hot radiation against
spontaneous condensation \cite{kapusta}. (This is altogether different from the
whole disappearance of the radiation by black hole accretion). During this
period both the viscosity coefficient and the relaxation time may be chosen as
monotonic decreasing functions, provided $\zeta/\tau$ grows with time.

Again we may interpret this behaviour in terms of a two-fluid model, where the
viscosity coefficient arises because of the particle production process from
the decay of massive non-relativistic particles into light ones. Shortly after
the beginning of the decay the particle production rate is large and the
energy density of the fluid becomes dominated by the light component. Later
on, as the decay rate slows down, the effect of adiabatic dilution by the fast
exponential expansion of the Universe turns out to be  more important.
Accordingly the non-relativistic energy density takes over again, since 
it goes down as $a^{-3}$, while the relativistic energy density goes 
down at the faster rate of $a^{-4}$.

The adiabatic index (\ref{gammat2}) can be used to estimate the dissipative
contribution to the speed of sound $v^2$ in (\ref{zeta})

\begin{equation}
\n{v}
v^2=\frac{2H_0^2\zeta}{\left(4\tau+3\zeta-2H_0\right)\left(H_0^2-
\frac{k}{a^2}\right)} \, .
\end{equation}
\noindent
In the limit $t\to\infty$, $v^2$ is a monotonic function of $\zeta$ and
$\tau$ within the region $2/3<\gamma$. It vanishes when $\zeta=0$
and reaches a maximum value $v^2=2/3$ when $\zeta/H_{0}\gg 1$.

\section{Concluding Remarks}
\noindent
We have applied the second method of Lyapunov to analyse the stability
of cosmic inflationary expansions, driven by a dissipative fluid governed
by a transport equation that allows for relaxation (i.e. of causal type),
for non spatially flat FLRW metrics. The parameters characterizing the
fluid (adiabatic index, viscosity coefficient and relaxation time) may
vary as the expansion proceeds. This is interesting as repeteadly 
stressed, a dissipative fluid can phenomenologically mimic a perfect 
one where particle production takes place \cite{Hu}, \cite{zeldovich}, 
\cite{barrow1}, \cite{winfried}, either from the quantum vacuum, or 
by the decay of pre-existing heavy  particles, or by the decay of 
massive modes of fundamental strings into massless modes.   

Essentially the curvature does not modify the results found in our previous
paper \cite{lad} concerning the stability of the de Sitter solution. This
result shows the insensibility of the viscosity-driven inflationary scenario
with respect to initial conditions even when spatial curvature is present.
However, it severely restrics the set of de Sitter solutions encompassed by
this model because a constraint (\ref{con}), linking the adiabatic index with
the relaxation time, must be satisfied. Nevertheless, since the de Sitter
solution is asymptotically stable for a wide set of behaviours of the fluid
parameters, inflation appears rather natural. The static solution, $ H = 0$,
is found to be unstable; however this result may be altered if a non-standard
increasing adiabatic index could enter the play. This work generalizes own
previous study in spatially-flat FLRW universes \cite{lad} as well as those of
Barrow \cite{barrow1}, \cite{barrow2}. Further, new solutions describing the
effects of dissipation have also been found. They either go over a stable de
Sitter scenario from an initial power--law singularity or asymptotically
approach a static universe if a fine tuning of the initial condition is made.

Lastly, an expression for the dissipative contribution to the
sound speed have been obtained (\ref{v}). In general the latter 
goes down as the Universe expands, something rather natural as the
impact of disipative effects is thought to diminish with expansion.

\nonumsection{Acknowledgments}
\noindent
This research has been partially supported by the Spanish Ministry
of Education under grant PB94-0718. LPC and ASJ thank the University
of Buenos Aires for partial support under project TX-93.

\nonumsection{References}

\nonumsection{Appendix}
\noindent
Equation (\ref{dH3}) with the change of variable  $a=s^{2/3\gamma}$ 
becomes
\begin{equation} \label{A1}
\ddot s+F(s)=0, \qquad F(s)=\frac{3\gamma}{2}
\left(\frac{3\gamma}{2}-1\right)k s^{1-4/3\gamma}-\frac{3\gamma c}{2}s.
\end{equation}

\noindent
Then, by the nonlocal transformation 

\begin{equation} \label{A2}
z=\int F(s) ds, \qquad \eta=\int F(s) dt
\end{equation}

\noindent equation (\ref{A1}) turns into a linear second order differential
equation with constant coefficient

\begin{equation} \label{A3}
z''+1=0
\end{equation}

\no where the prime indicates derivative with respect to $\eta$. Solving it 
the general solution of (\ref{A1}) in parametric form follows. This
can be achieved by inserting this solution in the second equation 
of (\ref{A2})

\begin{equation} \label{A4}
-\frac{\eta^2}{2}+c_1\eta+c_2=\frac{9\gamma^2}{8}k
s^{2\left(1-2/3\gamma\right)}-\frac{3\gamma c}{4}s^2
\end{equation}
\no The transformation law between $\eta$ and $t$ follows from solving
(\ref{A4}) for $s$ and inserting the resulting expression in the second
equation of (\ref{A2})

\begin{equation}
\n{A5}
t-t_0=\int \frac{d\eta}{F(s(\eta))} .
\end{equation}

\vskip 3cm
\noindent
{\Large {\bf Figure Caption}}

\vskip 1cm
\noindent
{\bf Figure 1} Graph of the potential $V(H)$, given by equation (\ref{V}), 
for each value of the spatial curvature $k$.
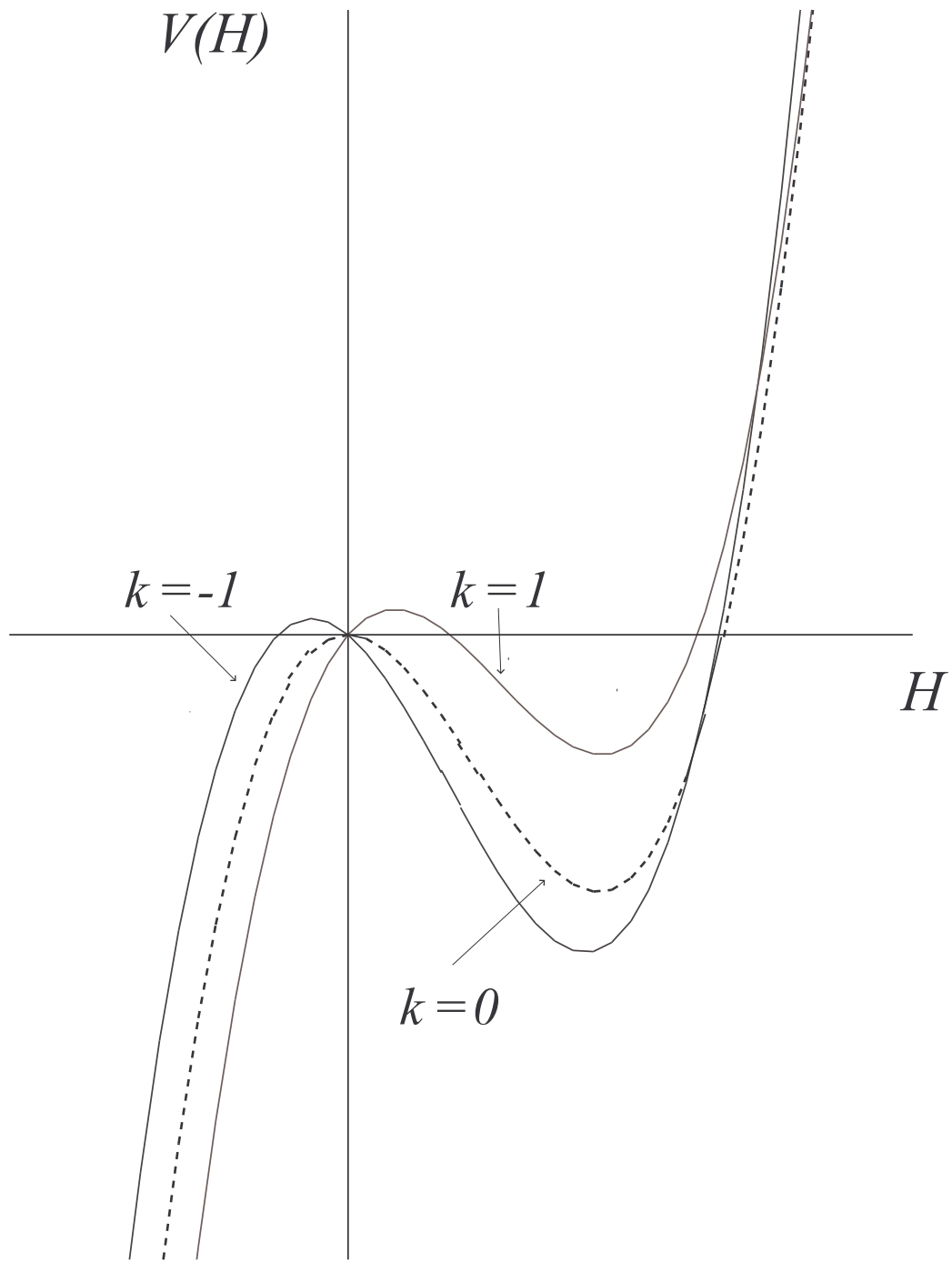
\end{document}